\documentclass[aps,prl,nofootinbib,reprint]{revtex4-1}

\usepackage{amsmath,amssymb,mathrsfs}

\begin{document}

\title{Black hole entropy from graviton entanglement}
\author{Eugenio~Bianchi}
\affiliation{Perimeter Institute for Theoretical Physics\\
31 Caroline St.N, Waterloo ON, N2J 2Y5, Canada}

\begin{abstract}
We argue that the entropy of a black hole is due to the entanglement of matter fields and gravitons across the horizon. While the entanglement entropy of the vacuum is divergent because of UV correlations, we show that low-energy perturbations of the vacuum result in a finite change in the entanglement entropy. The change is proportional to the energy flux through the horizon, and equals the change in area of the event horizon divided by $4$ times Newton's constant -- independently from the number and type of matter fields. The phenomenon is local in nature and applies both to black hole horizons and to cosmological horizons, thus providing a microscopic derivation of the Bekenstein-Hawking  area law. The physical mechanism presented relies on the universal coupling of gravitons to the energy-momentum tensor, i.e. on the equivalence principle. 
\end{abstract}

\maketitle

Black holes are perhaps the most perfectly thermal objects in the universe: there is convincing theoretical evidence that they are hot, they emit thermal radiation, and they are endowed with a thermodynamic entropy \cite{BH:73,Wald:1995yp}. A black hole can be perturbed away from its equilibrium state, for instance by letting some matter fall into it. When the system reaches equilibrium again, the area $A$ of the black hole horizon has grown by an amount $\delta A$. The increase in the entropy of the black hole $\delta S_{\text{\emph{BH}}}$ is given by the Bekenstein-Hawking formula,\footnote{We work with units $\kappa=c=\hbar=1$ and keep Newton's constant $G$ explicit. The metric signature is $(-+++)$.}
\begin{equation}
\delta S_{\text{\emph{BH}}}=\frac{\kappa\,c^3}{\hbar}\,\frac{\delta A}{4G}\,.\label{eq:S-BH}
\end{equation}
Explaining the microscopic origin of black hole's thermal nature is a long-standing problem in theoretical physics.

In this letter we prove that low-energy perturbations of the ground state of matter fields and gravitons on a black-hole background result in a variation of the horizon entanglement entropy $\delta S_{\text{ent}}$ that is finite, universal, and reproduces the Bekenstein-Hawking area law, $\delta S_{\text{ent}}=\delta S_{\text{\emph{BH}}}$. More specifically we consider a small patch that approximates Minkowski space at the horizon, and study the region bounded by a Rindler horizon. We show that an energy flux through the horizon corresponds to a variation of the entanglement entropy. This quantity is \emph{finite} and is shown to be independent of the number of matter \emph{species} because of the universal coupling of gravitons to the energy-momentum. We argue that this result establishes entanglement as the microscopic origin of black hole entropy \cite{EE:86}. The phenomenon is local in nature \cite{Jacobson:2003wv}, it holds for all non-extremal black holes \cite{BH:73}, and extends to the case of cosmological horizons \cite{Gibbons:1976ue}, thus explaining the universality of the formula (\ref{eq:S-BH}). 
The physical mechanism presented is a perturbative version of the one recently discovered in the context of loop quantum gravity where the entanglement entropy is finite because of non-perturbative effects \cite{Bianchi:2012fk}.

{\bf Vacuum entanglement}. The vacuum state of a quantum field has correlations at space-like separations. This entanglement implies that, although the vacuum is a pure state, the reduced density matrix associated to a region of space is mixed and there is an entanglement entropy $S_{\text{ent}}$ associated to it \cite{EE:86}. Such entropy is divergent and in 4 space-time dimensions scales as
\begin{equation}
S_{\text{ent}}=c_0 \,A_0\,\Lambda^2\,+c_1\log\Lambda\,+\,c_2\,,\label{eq:Sdiv}
\end{equation}
where $c_0$ and $c_1$ are numerical constants, $A_0$ is the area of the surface bounding the region, $\Lambda$ is a high-energy cut-off, and $c_2$ is the finite part of the entanglement entropy.

The entanglement entropy of field theoretical modes across the horizon has been extensively studied as a proposed explanation of black hole entropy \cite{EE:86}. As originally formulated however, the proposal suffered from three problems: (i) the entropy $S_{\text{ent}}$ in eq.(\ref{eq:Sdiv}) is divergent, (ii) to reproduce the $1/4$ prefactor of eq.(\ref{eq:S-BH}) the high energy cut-off $\Lambda$ has to be tuned at the Planck scale, (iii) such tuning depends on all the physics from low-energies up to the Planck scale, in particular on the number of species of matter fields we unfreeze going to higher and higher energies \cite{suss}. Here we present a resolution of these puzzles by considering low-energy processes in which the entanglement entropy changes. In such processes only the low-energy part of the entanglement entropy is perturbed, and the change $\delta S_{\text{ent}}$ is shown to be insensitive to the UV behavior of the entanglement entropy.

{\bf Perturbative quantum gravity.}  In perturbative quantum gravity \cite{Feynman:1996kb}, the graviton field $h_{\mu\nu}$ is defined as the perturbation of the space-time metric $g_{\mu\nu}$ about the flat Minkowski background $\eta_{\mu\nu}$,
\begin{equation}\label{eq:graviton}
g_{\mu\nu}=\eta_{\mu\nu}+\sqrt{32\pi G}\,h_{\mu\nu}\,.
\end{equation}
At leading order in a perturbative expansion in Newton's constant $G$, the action for gravity coupled to a scalar field is given by the Fierz-Pauli action for free gravitons, 
plus the action of a free scalar field $\phi$, plus a universal coupling of the graviton to the total energy-momentum tensor. Schematically,
\begin{equation}
I=I_\text{grav}[h_{\mu\nu}]+I_{\text{matt}}[\phi]+\sqrt{8\pi G}\int d^4x \,h_{\mu\nu} T^{\mu\nu}\,,\label{eq:action}
\end{equation}
where $T_{\mu\nu}$ is the energy-momentum of matter fields \emph{and} gravitons. Varying the action with respect to the graviton field, we find the equation of motion
\begin{equation}\label{eq:waveq}
\Box h_{\mu\nu}=-\sqrt{8\pi G}\; (T_{\mu\nu}-\frac{1}{2}\eta_{\mu\nu} T^\sigma_\sigma)\,,
\end{equation}
where $\Box=\partial_\mu \partial^\mu$ and we have adopted the harmonic gauge $\partial^\nu h_{\mu\nu}=\frac{1}{2}\partial_\mu {h^\nu}_{\nu}$. This equation will be crucial to the following.

{\bf The perturbed Rindler horizon.}  A uniformly accelerated observer in Minkowski space cannot receive signals from behind the Rindler horizon. Let $x^\mu=(t,x,y,z)$ be Cartesian coordinates adapted to an inertial observer, and $(\eta,\rho,y,z)$ be Rindler coordinates, where $\eta$ is a dimensionless ``angular'' coordinate, i.e. $t=\rho \sinh \eta$ and $x=\rho \cosh \eta$. The Minkowski line element is given by
\begin{equation}
ds^2=-\rho^2 d\eta^2+d\rho^2+dy^2+dz^2\,.
\end{equation}
An observer at $\rho=\ell$ has a uniform acceleration $a=\ell^{-1}$, and is causally connected only with the portion $x>t$ of Minkowski space. Its boundary is the Rindler horizon. Let $(v,y,z)$, with $v=t+x>0$, be coordinates on the future Rindler horizon $H$. We call $A_v$ the area of a small patch $\mathcal{B}_v$ of the space-like surface $(y,z)$ at given $v$.

In the presence of space-time curvature there is geodesic deviation. Consider the small perturbations $h_{\mu\nu}$ of the metric away from flat space, and a beam of light rays in $H$ that crosses the plane $v=0$ drawing on it the surface $\mathcal{B}_{0}$ of area $A_0$. We call  $k^\mu=\frac{\partial x^\mu}{\partial v}$ the tangent vectors to such light rays. At $v>0$ they will draw a new patch $\mathcal{B}_v$ with a different area. Because of the gravitational perturbation, the beam is focused. The event horizon, or \emph{perturbed} Rindler horizon,  is defined by light rays that are finely balanced between falling out of sight of the accelerated observer and escaping to infinity \cite{Wald:1995yp}. The area $A_H$ of the event horizon is defined as the area of the cross-section $\mathcal{B}_{\infty}$ of such light rays and given by
\begin{equation}
A_H\,=\,A_0+\sqrt{8\pi G}\int_0^{\infty}\!\!\int_{\mathcal{B}_v}\!\!\!(-\Box h_{\mu\nu}\;k^\mu k^\nu)\;v dv\,dydz\,,\label{eq:expansion}
\end{equation} 
at the leading order in $\sqrt{8\pi G}$. The effect is due to light deflection by a gravitational perturbation and can be described geometrically in terms of the Raychaudhuri equation for the expansion of null geodesics\footnote{In the harmonic gauge and at the linear order in the graviton field, the Ricci tensor is $R_{\mu\nu}=\nolinebreak-\sqrt{8\pi G}\,\Box h_{\mu\nu}$, and the expansion $\theta=\nabla_\mu k^\mu$ of null geodesics satisfies the linearized Raychaudhuri equation $\frac{\partial \theta}{\partial v}=\nolinebreak\sqrt{8\pi G}\; \Box h_{\mu\nu}\, k^\mu k^\nu$. The event horizon has expansion $\theta(v)=\sqrt{8\pi G}\int^{\infty}_v - \Box h_{\mu\nu}\, k^\mu k^\nu\,dv'$. This is the advanced solution of the Raychaudhuri equation with final boundary condition $\theta(\infty)=0$. The area change is related to the expansion by the equation $\Delta A=\int\theta\, dv\,dydz$.  See \cite{Jacobson:2003wv} for a detailed analysis.} \cite{Wald:1995yp}.

{\bf Entanglement and thermality.} The Minkowski vacuum state $|\Omega\rangle$ of the interacting quantum field theory described by the action (\ref{eq:action}) is Poincar\'e invariant. In particular it is invariant under boosts in the $x$ direction and therefore it appears stationary to the uniformly accelerated observers discussed above. Moreover, being the ground state, it is stable under dynamical perturbations. These two properties suggest that accelerated observers see the Minkowski vacuum as a thermal equilibrium state. That this is in fact the case is the core of the Unruh effect \cite{Unruh:1976db}.  The physical mechanism behind its thermality is the entanglement with modes across the Rindler horizon. This fact is manifest when the interacting Minkowski vacuum of matter fields and gravitons is written in the form \cite{Wald:1995yp}
\begin{equation}
|\Omega\rangle=\frac{1}{\sqrt{Z}}\int D\varphi_L D\varphi_R\;\langle\varphi_L|e^{-\pi K}|\varphi_R\rangle\;   |\varphi_L\rangle\otimes|\varphi_R\rangle\,,
\end{equation}
where $\varphi_L$ stands for the matter field and the graviton $(\phi,h_{\mu\nu})$ both restricted to $x<0$, that is to the \emph{left} half-space, and similarly $\varphi_R$ for the \emph{right} half-space\footnote{This property of the Minkowski vacuum can be derived using a path integral representation for the ground state of an interacting theory \cite{Wald:1995yp,Unruh:1976db}. It generalizes to the Hartle-Hawking vacuum of quantum fields on a static black-hole background \cite{Hartle:1976tp} and on de Sitter space \cite{Gibbons:1976ue}, where similar cross-horizon correlations appear.}. The hermitian operator $K$ is the Rindler Hamiltonian
\begin{equation}
K=\int T_{\mu\nu}\chi^\mu\,d\Sigma^\nu\,.
\end{equation}
where $\chi^\mu=\frac{\partial x^\mu}{\partial \eta}$ is the boost Killing vector, $\xi^\mu=\frac{1}{\rho}\frac{\partial x^\mu}{\partial \eta}$ the same vector normalized to $-1$, and $d\Sigma^\mu=\xi^\mu\, d\rho\, dydz$ is the volume element. The operator $K$ is the generator of boosts, i.e. translations in $\eta$. As accelerated observers cannot probe the region $x<0$, all their observations can be described using the reduced density matrix
\begin{equation}
\rho_0=\text{Tr}_L |\Omega\rangle\langle\Omega|\;=\;\frac{e^{-2\pi K}}{Z}\;,\label{eq:gibbs}
\end{equation}
where the trace is on left modes $\varphi_L$ of the matter field and the graviton. This density matrix represents a Gibbs state \cite{Bisognano:1975ih}, it is thermal with respect to boost evolution with associated geometric\footnote{The temperature $T_{\text{geom}}$ is dimensionless. An observer with acceleration $a$ and tangent vector $u^\mu=a\,\partial x^\mu/\partial \eta$  measures an energy density $T_{\mu\nu} u^\mu \xi^\nu$ and a  local temperature $T_{\text{loc}}=a/2\pi$, the Unruh temperature.} temperature $T_{\text{geom}}=\frac{1}{2\pi}$, \cite{Unruh:1976db}.  The horizon is \emph{hot} because of entanglement.

{\bf Area law from entanglement perturbations.}  The entanglement entropy of the Minkowski vacuum $|\Omega\rangle$ is defined as the von Neumann entropy of its reduced density matrix $S_{\text{ent}}(|\Omega\rangle)=-\text{Tr}(\rho_0 \log \rho_0)$. This expression is divergent due to the high-energy correlations present in the vacuum state as explained above. Notice however that in thermodynamics the physically relevant quantity is not the entropy in itself, but how it changes in a physical process. We consider now a perturbation of the vacuum corresponding to an energy flux through the Rindler horizon, and compute the variation of the entanglement entropy during the process.

Let $|\mathcal{E}\rangle$ be an excited state of the system (\ref{eq:action}) of interacting gravitons and matter fields. It could be for instance the state of a particle crossing the Rindler horizon \cite{Marolf:2003sq}. We call $\rho_1=\text{Tr}_L|\mathcal{E}\rangle\langle \mathcal{E}|$ its reduced density matrix. If the state has low energy, then $\rho_1$ can be considered as a small perturbation of $\rho_0$, i.e. $\rho_1=\rho_0+\delta \rho$. The entanglement entropy $S_{\text{ent}}(|\mathcal{E}\rangle)= -\text{Tr}(\rho_1 \log \rho_1)$ has exactly the same UV structure (\ref{eq:Sdiv}) as the vacuum state because only the low-energy modes of the two differ. In particular, the change in entropy  
\begin{equation}
\delta S_{\text{ent}}= S_{\text{ent}}(|\mathcal{E}\rangle)-S_{\text{ent}}(|\Omega\rangle)
\end{equation}
is finite and can easily be computed at the first order in the perturbation $\delta \rho$,
\begin{equation}
\delta S_{\text{ent}}=-\text{Tr} (\delta\rho\, \log \rho_0) =\,2\pi\, \text{Tr}(K\,\delta \rho)\,.\label{eq:dS=2pidE}
\end{equation}
The first equality follows from the identity $\text{Tr}(\delta \rho)=0$, the second from the Gibbs form (\ref{eq:gibbs}) of $\rho_0$. Moreover, as the boost energy is conserved, we can evaluate it in the limit of large $\eta$ in which the vector $\xi^\mu$ becomes light-like and parallel to $k^\mu$. In this limit, if no energy escapes to future null infinity\footnote{The energy flux at future null infinity vanishes for a generic state because the only way for a wave not to fall across the Rindler horizon is to travel in the direction exactly perpendicular to the horizon as argued in \cite{Wall:2010cj}.}, the boost energy becomes an integral over the horizon,  
\begin{equation}
K=\int_H T_{\mu\nu}\,v\, k^\mu\,dH^\nu\,,
\end{equation}
where $k^\mu=\frac{\partial x^\mu}{\partial v}$, $dH^\mu=k^\mu dv \,dydz$, and we have used\footnote{We use the symbol $\stackrel{\;H\!\!}{=}$ to denote equality on the horizon $H$.} $\chi^\mu\stackrel{\;H\!\!}{=}v k^\mu$, and $d\Sigma^\mu\stackrel{\;H\!\!}{=}dH^\mu$. As a result
\begin{equation}
\delta S_{\text{ent}}\,=\,2\pi\; \text{Tr}\big(\!\int_H T_{\mu\nu}k^\mu k^\nu\,vdv\,dydz\;\,\delta \rho\big)\,.\label{eq:S=intH}
\end{equation}
We can now express the energy-momentum of matter and gravitons in terms of the d'Alembertian of the graviton field using the wave equation (\ref{eq:waveq}),
\begin{equation}
\delta S_{\text{ent}}\,=\,\frac{2\pi }{\sqrt{8\pi G}}\text{Tr}\big(\!\int_H\!\!-\Box h_{\mu\nu}\,k^\mu k^\nu\,vdv\,dydz\;\,\delta \rho\big)\,.
\end{equation}
As $\Box h_{\mu\nu}$ controls the deflection of light rays and the expansion of the area of the surface $\mathcal{B}_v$, we have
\begin{equation}
\delta S_{\text{ent}}\,=\,2\pi \frac{1}{8\pi G}\text{Tr}(A_H\, \delta \rho)\;=\;\frac{\delta A}{4G}\,,\label{eq:dSent}
\end{equation}
where $A_H$ is the operator defined in terms of the graviton field $h_{\mu\nu}$ and the background value $A_0$ by eq.(\ref{eq:expansion}), and $\delta A$ is the change in the area of the event horizon. The result reproduces the variation of the Bekenstein-Hawking entropy \cite{BH:73} of the Rindler horizon \cite{Jacobson:2003wv}. The entropy is due to the variation in the amount of entanglement for matter fields \emph{and} gravitons, and has a universal expression in terms of the area change $\delta A$ thanks to the universal coupling of gravitons to the energy-momentum tensor.

{\bf Black hole horizons and cosmological horizons}. The result presented above for the Rindler horizon applies directly to the entropy of black hole horizons \cite{BH:73} and of cosmological horizons \cite{Gibbons:1976ue}. The argument hinges on the local nature of the results presented: we considered a small surface $\mathcal{B}_0$ and associated to it an entanglement entropy per unit surface area. As discussed in \cite{Jacobson:2003wv}, this \emph{entropy density} governs near-horizon thermodynamic processes for all horizons. More specifically, in the case of a non-extremal Kerr-Newman black hole, we can consider a near-horizon co-rotating frame. The local geometry is stationary and described by the Rindler metric\footnote{Near-horizon stationary observers have acceleration $a=\ell^{-1}$ and coincide with the ZAMOs of \cite{Thorne:1986iy} as $\ell\to 0$. Photons at the local Unruh temperature $T_{\text{loc}}=a/2\pi$ that manage to escape to infinity will be red-shifted to the Hawking temperature \cite{BH:73}}. Compatibly with the equivalence principle, the vacuum state behaves locally as the Minkowski vacuum \cite{frolov1989renormalized,Hollands:2008vx}, and we can apply our analysis to this system: a perturbation of the vacuum corresponding to an energy-flux through the horizon results in an increase of the horizon entanglement entropy by $\delta S_{\text{ent}}=\delta A/4G$, where $\delta A$ is the area change of the event horizon of the black hole. The condition of applicability of our approximation is that the perturbation is of low energy with respect to the scale of the UV cut-off and of short wavelength compared to curvature scale at the black-hole horizon.

Similarly in the case of de Sitter space, we can consider a static patch. Static observers near the cosmological horizon have a large accelaration $a=\ell^{-1}$, where $\ell$ is the proper distance from the horizon.  The near-horizon geometry is again Rindler, independently of the positive value of the cosmological constant. Near-horizon processes can be considered \cite{davies1984mining} and the result (\ref{eq:dSent}) applies: the change in the entanglement entropy is $\delta A/4G$ where $\delta A$ is the increase in area of the cosmological horizon.

{\bf Conclusions.} In this letter we have shown that entanglement provides the microscopic explanation of black hole entropy: causal horizons have thermal properties because of entanglement, and the Bekenstein-Hawking formula describes the variation in the amount of entanglement during perturbations that have low-energies compared to the scale of a physical cut-off $\Lambda$. Such pertubations change only the infrared part of the entanglement entropy, leaving the ultraviolet part untouched. The variation $\delta S_{\text{ent}}$ is finite and universal because of gravitational backreaction: the universal coupling of gravitons to all matter allows one to express the energy flux through the horizon in terms of the deflection of the light rays that define the horizon, therefore reproducing the geometric quantity $\delta A/4G$. The area $A_0$ of the small patch $\mathcal{B}_0$ appearing in eq.(\ref{eq:Sdiv}) is unchanged, what changes is the area $A_H$ of the event horizon defined by eq.(\ref{eq:expansion}). The phenomenon persists even in the absence of matter fields as gravitons couple to their own energy-momentum tensor. The derivation of the universality of $\delta S_{\text{ent}}$ rests on the equivalence principle, here formulated in a particle physicist's language \cite{Feynman:1996kb}.


It is often argued that a microscopic derivation of black hole entropy should consist in a counting of microstates \cite{bh:string,bh:lqg}. The explanation of the horizon entropy in terms of entanglement of gravitons is compatible with the (apparently alternative) proposal that the Bekenstein-Hawking entropy is due to thermal fluctuations of the shape of the horizon \cite{York:1983zb}. In fact, the entangled vacuum state $|\Omega\rangle$ results in a reduced density matrix $\rho_0$ that is thermal with respect to the flow generated by the boost Hamiltonian. Therefore the system can be described pertubatively as a gas of gravitons interacting with matter fields and thermalized by a \emph{membrane} at $l=\Lambda^{-1}$ with temperature $T_0=\frac{\Lambda}{2\pi}$ \cite{Hooft:1984re,Thorne:1986iy}. Thermal gravitons correspond to thermal fluctuations in the shape of the horizon. Counting states for the thermal system \`a la Boltzmann results in an entropy that is divergent with the UV cut-off $\Lambda$. However low-energy perturbations of the thermal state result in finite variations $\delta S_{\text{therm}}=\delta S_{\text{ent}}$ in the entropy. Such variations are universal and equal to the variation $\delta A/4G$ in the area of the horizon, thanks to the universal coupling of gravitons to the energy-momentum tensor. From the perspective of the outside observer, we cannot distinguish thermal fluctuations from quantum cross-horizon correlations \cite{smolin1999nature}. What the thermal picture cannot account for is why the membrane is hot. The answer is because of entanglement.

In loop quantum gravity the entanglement of quantum geometries across the horizon is finite because the theory has no degrees of freedom above the energy scale $\gamma^{-1/2} \Lambda_{Planck}$, where $\gamma$ is the Immirzi parameter. While the entropy diverges if $\gamma$ is sent to zero, the variation $\delta S_{\text{ent}}$ corresponding to a change in the horizon energy is independent of $\gamma$ and reproduces the Bekenstein-Hawking formula by the same physical mechanism discussed above \cite{Bianchi:2012fk}. The availability of a non-perturbative framework opens up the intriguing possibility for answering questions as: ``What is the fate of the divergent part of the entanglement entropy? How can we probe it?'' Clearly, this term can contribute to thermal properties of the system only if it can be changed in a physical process. For instance, for photons the typical scale of the physical cut-off is $1\, \text{MeV}$, which is when electron-positron pairs are produced. When this threshold is reached a new contribution to the entropy appears. Similarly for other matter fields, so that we may argue that at high energies there is an extra chemical-potential contribution $\mu$ to the Bekenstein-Hawking entropy $\delta S=2\pi(\frac{\delta A}{8\pi G}\,+\,\mu\,\delta N)$ where $N$ is the number of unfrozen matter fields. An analogous phenomenon is being investigated in loop quantum gravity \cite{Ghosh:2011fc} and, together with the derivation \cite{Bianchi:2012fk} of the horizon entanglement entropy, it may lead to new physical effects associated to the high-energy behavior of causal horizons.

\vspace{.5em}

\emph{Acknowledgements}. Thanks to F.~Cachazo, L.~Freidel, T.~Jacobson, R.~Myers, C.~Rovelli, R.~Sorkin, and A.~Satz  for discussions. This research is supported by the Perimeter Institute for Theoretical Physics and by a Banting Fellowship through NSERC.


\providecommand{\href}[2]{#2}\begingroup\raggedright\endgroup

\end{document}